\documentclass[12pt]{iopart}

\usepackage{iopams}  
\usepackage{graphicx}
\usepackage{epstopdf}
\begin{document}

\title[The role of angular momentum in the construction of multipolar fields]{\bf{The role of angular momentum in the construction of electromagnetic multipolar fields}}

\author{Nora Tischler, Xavier Zambrana-Puyalto and Gabriel Molina-Terriza}

\address{QISS and Department of Physics \& Astronomy, Macquarie University, 2109 NSW, Australia}
\address{ARC Centre of Excellence for Engineered Quantum Systems}
\ead{gabriel.molina-terriza@mq.edu.au}

\date{\today}

\begin{abstract}
Multipolar solutions of Maxwell's equations are used in many practical applications and are essential for the understanding of light-matter interactions at the fundamental level. Unlike the set of plane wave solutions of electromagnetic fields, the multipolar solutions do not share a standard derivation or notation. As a result, expressions originating from different derivations can be difficult to compare. Some of the derivations of the multipolar solutions do not explicitly show their relation to the angular momentum operators, thus hiding important properties of these solutions. In this article, the relation between two of the most common derivations of this set of solutions is explicitly shown and their relation to the angular momentum operators is exposed.   
\end{abstract}

\maketitle       

\section{Introduction}\label{Introduction}
The solution to Maxwell's equations in homogeneous isotropic media is one of the most studied problems in undergraduate physics. Often the problem is solved in Cartesian coordinates and gives rise to the typical plane wave solutions of electromagnetic fields. Although other solutions in different coordinate systems are not so widespread, the multipolar fields arising from the use of spherical coordinates stand out due to their use in several important problems in electromagnetism \cite{Jackson}. These sets of electromagnetic fields play a very important role in the interaction of light with matter. In particular, they are widely used to understand the atomic and molecular spectra \cite{Craig_Thiru,Blatt_Weisskopf}. More recently, the multipolar fields have found interesting applications in the field of nanophotonics \cite{Novotny}. Probably the most successful application of this set of solutions is the Mie scattering problem, which studies the interaction of electromagnetic plane waves with spherical particles \cite{Mie}. Mie theory provides one of the few analytical solutions of the inhomogeneous Maxwell's equations. This theory generalizes Rayleigh scattering, which accounts for the colour of the sky, to particles of any size. For all these reasons, the multipolar fields appear in almost all undergraduate textbooks and in a long range of specialized books. Unfortunately, the multipolar fields do not share the structural simplicity of the plane wave solution. This fact translates into a myriad of different representations for the multipolar solutions. Different books use notations and derivations which are so dissimilar that the study of these solutions can be difficult. More importantly, some of the standard derivations hide the beautiful symmetry of the multipolar fields and can prevent a deeper understanding of the underlying physics, i.e. the angular momentum of light. 

In this article we would like to present the most common derivations of the multipolar fields and how the different notations are related. This exercise usually requires a significant amount of algebra and is rather cumbersome \cite{Bouwkamp}. Here, we will show that an insightful understanding of the role of the angular momentum operator in electromagnetism and the proper use of some of its properties significantly simplifies the calculations. With this new understanding, the multipolar solutions become much clearer and can be used in physical problems with a new point of view.

The remainder of the article is organized as follows: In section \ref{Multipolar fields} the problem is outlined, along with two common derivations of the multipolar solutions. By making use of the orbital angular momentum operator, the correspondence between these two sets of solutions is discussed in section \ref{Correspondence}. Finally, section \ref{Conclusion} concludes the work with a summary and outlook.

\section{Multipolar fields}\label{Multipolar fields}
Let us start by reviewing the multipolar solutions to Maxwell's equations for source-free homogeneous and isotropic media. We consider monochromatic electric, $\mathcal{E}$, and magnetic, $\mathcal{H}$, fields, i.e. $\mathcal{E}=\Re e\{\mathbf{E} \exp(-i \omega t)\}$ and similarly for the magnetic field $\mathcal{H}=\Re e\{\mathbf{H} \exp(-i \omega t)\}$, where the vectors $\mathbf{E}$ and $\mathbf{H}$ follow Maxwell's equations and therefore the resulting vector wave equations
\begin{equation}
\nabla^2\mathbf{E} + k^2 \mathbf{E} = \mathbf{0}, ~~~~~ \nabla^2\mathbf{H} + k^2\mathbf{H} = \mathbf{0}, 
\label{Maxwell1}
\end{equation}
with $k=\sqrt{\epsilon \mu} \omega$, have zero divergence
\begin{equation}
\nabla \cdot \mathbf{E} = 0,  ~~~~~ \nabla \cdot \mathbf{H} = 0,
\label{Maxwell2}
\end{equation}
and are interrelated:
\begin{equation}
\nabla \times \mathbf{E} = i \omega \mu \mathbf{H},  ~~~~~ \nabla \times \mathbf{H} = -i\omega\epsilon\mathbf{E}.
\label{Maxwell3}
\end{equation}
Here $\epsilon=\epsilon_0 \epsilon_r$, where $\epsilon_0$ and $\epsilon_r$ are the vacuum and relative permittivity, respectively, and similarly $\mu=\mu_0 \mu_r$, where $\mu_0$ and $\mu_r$ are the vacuum and relative permeability.\\
Multipolar fields are solutions of the above equations in spherical coordinates. They may be chosen to be simultaneously eigenfunctions of the squared angular momentum of the radiation field, a component of the angular momentum on one of the coordinate axes (here, the z-axis is chosen), and of the parity operator. The multipolar fields are often used as basis functions, into which electromagnetic fields can be decomposed \cite{Bohren}. Specification of the weights in such a multipolar expansion allows the description of electromagnetic fields.

\subsection{Derivation of the multipolar solutions of Maxwell's equations}

One can find several expressions for the multipolar solutions of Maxwell's equations in the literature, but they are typically derived in two different ways. The first one uses a well-known recipe for deriving the solution of (\ref{Maxwell1}),(\ref{Maxwell2}) in a medium free of charges \cite{Morse}. This recipe is applicable to different coordinate systems that meet certain conditions \cite{Morse}, and consists in using the following fields:
\begin{equation}
\mathbf{M} \equiv \nabla \times (\mathbf{u}\tau), ~~~~~\mathbf{N} \equiv \frac{\nabla \times \mathbf{M}}{k},
\label{general}
\end{equation} 
where $\mathbf{u}$ is a unit vector in the symmetric direction and $\tau$ is a scalar function expressed in the appropiate system of coordinates. This scalar function $\tau$ can be written as $\tau=w(\chi)\psi$, where $\chi$ is the coordinate associated with the available symmetry and $w(\chi)=\{1,\chi \}$ depending on the problem; $\psi$ is a solution of the scalar Helmholtz equation in the chosen system of coordinates. If all these conditions are fulfilled, the vector fields \textbf{M} and \textbf{N} are solutions of Maxwell's equations showing the appropriate symmetries. 

In the case of the multipolar solutions, the system of coordinates is spherical, the unit vector points in the radial direction, $\mathbf{u}=\mathbf{\hat{r}}$, and $w(r)=r$. This approach is typically found in Mie theory related literature \cite{Bohren} as well as in some electromagnetism textbooks \cite{Novotny,Reitz,Stratton}. We will refer to this method as the ``general method,'' since it may also be used to obtain solutions in other geometries.

Another way of deriving the multipolar solutions of Maxwell's equations uses the fact that these solutions have to be eigenvectors of the angular momentum operator. Hence, the solution can be built from the rules of addition of the angular momentum operator. This construction is detailed in the books by Rose \cite{Rose_book,Rose_booklet}. In this method, the multipole solutions are explicitly shown to be eigenvectors of the angular momentum operator, a fact that is typically overlooked in many books about Mie theory. We will refer to this method as the ``angular momentum method,'' as it explicitly makes use of the angular momentum operator properties.

The diametrically different construction methods of the multipole solutions may lead to some confusion due to several factors. On the one hand, as we will see immediately, both methods naturally give rise to different notations, which are rather cumbersome to reconcile \cite{Bouwkamp}. Then, one may be tempted to think that the different multipole notations belong to different classes of solutions when in reality they represent the same object. On the other hand, the derivation of the multipolar solutions without making use of their role as eigenvectors of the angular momentum operators hides some important properties of this set of solutions. For example, it is often difficult to understand the relation between the different resonances in Mie scattering and the angular momentum of the field and/or the particle. In this paper, we will try to solve this problem once and for all and show how the two construction methods are related. 

\subsection{General method}

As mentioned before, the general method is based on (\ref{general}). The scalar function $\tau$ is one of the solutions of the scalar Helmholtz equation in spherical coordinates multiplied by the radial coordinate $(\tau=r\psi)$, and the unit vector is oriented along the radial direction $(\mathbf{u}=\mathbf{\hat{r}})$. Our notation closely follows that of Bohren and Huffman \cite{Bohren}, which is very common in Mie theory books. Note, however, that we distinguish different modes with the indices $M$ and $L$, instead of the $m$ and $n$ that are used in the reference. The choice for this trivial change will become clear in the following sections. Using (\ref{general}), we define a set of solutions $\{\mathbf{M}_{eML}, \mathbf{M}_{oML}, \mathbf{N}_{eML}, \mathbf{N}_{oML}\}$, which are determined by the scalar functions $\psi_{eML} = \cos(M\phi)P_L^M(\cos(\theta))z_L(kr)$ and $\psi_{oML} = \sin(M\phi)P_L^M(\cos(\theta))z_L(kr)$. Here, as in the rest of the article, $(r,\theta,\phi)$ are the spherical coordinates ($\phi$ being the azimuthal angle). The radial function $z_L(kr)$ can be either a spherical Bessel function or spherical Hankel function. Within the general method the indices $M,L$ are non-negative integers and $L \ge M$. As required by the recipe of (\ref{general}), the functions $\psi_{oML}$ and $\psi_{eML}$ fulfill the scalar Helmholtz equation in spherical coordinates. The definitions required to construct the complete form of the solutions $\{\mathbf{M}_{eML}, \mathbf{M}_{oML}, \mathbf{N}_{eML}, \mathbf{N}_{oML}\}$ can be found in the appendix. As an example, 
\begin{equation}\label{Mexample}
\fl \mathbf{M}_{oML} = \frac{M}{\sin (\theta)}\cos (M\phi) P_L^M (\cos (\theta)) z_L (kr) \hat{\mathbf{e}}_{\theta}-\sin(M\phi)\frac{dP_L^M (\cos (\theta))}{d\theta}z_L(kr) \hat{\mathbf{e}}_\phi,
\end{equation}
where $\hat{\mathbf{e}}_{\theta}$ and $\hat{\mathbf{e}}_{\phi}$ are the polar and azimuthal unit vectors.

Note that there are two different sets of ${\mathbf{M}}$ and ${\mathbf{N}}$ solutions labeled with either $o$ (odd) or $e$ (even), and they differ in the azimuthal behaviour of the scalar function $\psi$, oscillating as a sine or a cosine, respectively. The main difference between these multipolar fields and those given by Jackson \cite{Jackson} is that instead of sinusoidal functions, the author uses complex exponential functions.  

\subsection{Angular momentum method} \label{ang_mom_method_derivation}

As mentioned earlier, the general method does not make use of the fact that the multipolar solutions should have a well defined relation with the angular momentum operator.\footnote{Of course, the angular momentum content of the multipolar solutions can always be calculated and gives the expected result \cite{Jackson}.} However, one can derive the multipolar solutions by directly imposing that they should be eigenvectors of the angular momentum operator. We will follow the notation by Rose \cite{Rose_book,Rose_booklet}. First we start by noting that the angular momentum operator ($\mathbf{J}$) for a vectorial field is composed of two terms: the orbital angular momentum $\mathbf{L}=-i( \mathbf{r}\times\nabla)$ and the spin angular momentum $\mathbf{S}$, i.e. $\mathbf{J}=\mathbf{L}+\mathbf{S}$. In this article, we use dimensionless angular momentum operators, as in the references cited \cite{Jackson,Rose_book,Rose_booklet}. The angular momentum operator so defined is the generator of rotations for the vectorial electromagnetic field. It is also related to the density of electromagnetic angular momentum of the field (i.e. $ \mathbf{m}=\Re e \{ \mathbf{r} \times \mathbf{E} \times \mathbf{H}^\ast\}$). The interested reader can find more details about these subtleties in the given literature.

We should now impose that the multipolar vector fields are eigenvectors of $\mathbf{J}^2$ and $J_z$. We can start constructing these solutions by composing the eigenfunctions of both $\mathbf{L}$ and $\mathbf{S}$. The eigenfunctions of $\mathbf{L}^2$ and $L_z$ are the spherical harmonic scalar functions ($Y_L^M(\theta,\phi)$):
\begin{equation}
\mathbf{L}^2 Y_L^M=L (L+1) Y_L^M,~~~~~ \,\, L_z Y_L^M =M Y_L^M.
\end{equation}
Note that the above equations can be considered as the definition for the spherical harmonics as they are a set of partial differential equations. On the other hand, the eigenfunctions of $\mathbf{S}^2$ and $S_z$ are the vectors $\boldsymbol{\xi}_{+1}=-(\hat{\mathbf{x}}+ i \hat{\mathbf{y}})/\sqrt{2}$, $\boldsymbol{\xi}_{0}=\hat{\mathbf{z}}$, and $\boldsymbol{\xi}_{-1}=(\hat{\mathbf{x}}- i \hat{\mathbf{y}})/\sqrt{2}$:
\begin{equation}
\mathbf{S}^2 \boldsymbol{\xi}_\mu=2 \boldsymbol{\xi}_\mu,~~~~ \,\, S_z \boldsymbol{\xi}_\mu =\mu \boldsymbol{\xi}_\mu.
\end{equation}

We then construct the multipolar vector functions $\mathbf{T}_{JLM}$ by combining the scalar eigenfunctions ($Y_L^{M+\mu}$) and the vector eigenfunctions ($\boldsymbol{\xi}_{-\mu}$), using the  rules of angular momentum addition. A simple product of the scalar and vector eigenfunctions, $Y_L^{M+\mu} \boldsymbol{\xi}_{-\mu}$, remains an eigenfunction of the orbital and spin angular momentum operators. However, it does not have the required properties with regard to the squared total angular momentum operator, $\mathbf{J}^2$. In order to create eigenfunctions of the squared total angular momentum operator, one has to combine the separate eigenfunctions with the aid of the Clebsch-Gordan coefficients:
\begin{equation}
\mathbf{T}_{JLM}=\sum_{\mu=-1}^{\mu=+1} C(1LJ;-\mu,M+\mu) Y_L^{M+\mu}(\theta,\phi)\boldsymbol{\xi}_{-\mu} \label{TJLM},
\end{equation}
where $C(abc;d,e)$ denotes the well-known Clebsch-Gordan coefficients, which we have explicitly written in the appendix. Note that $\mathbf{T}_{JLM}$  is an eigenfunction of $J_z$, because the three terms in the sum above ($\mu=-1,0,+1$) combine eigenfunctions with $L_z=M+\mu$ and $S_z=-\mu$, always summing $J_z=L_z+S_z=M$. Hence, in this notation M has become the eigenvalue of the z-component of the total angular momentum operator. Furthermore, the weights of the three components in the sum are the Clebsch-Gordan coefficients, which are exactly calculated so that the total angular momentum fulfills $\mathbf{J}^2\mathbf{T}_{JLM}=J(J+1)\mathbf{T}_{JLM}$. In summary, the vector functions $\mathbf{T}_{JLM}$ are eigenvectors of both $\mathbf{J}^2$ and $J_z$, but also of $\mathbf{L}^2$ and $\mathbf{S}^2$. They are often referred to as vector spherical harmonics and are also used by Jackson \cite{Jackson}, even though the derivation in this reference is different. A further noteworthy property of these vector spherical harmonics is that they are orthonormal on the surface of a unit sphere.

The electric and magnetic multipolar solutions can be derived from the vector spherical harmonics in (\ref{TJLM}). The addition rules for angular momenta impose that for a fixed $J$, the orbital angular momentum $L$ can only take three values: $J-1$, $J$, and $J+1$. To proceed, we can make use of the fact that the multipolar fields are also required to be eigenvectors of the parity operator, P. This means that they are invariant under point inversions $(\mathbf{r}\rightarrow -\mathbf{r})$, except for possibly an overall phase difference.  
The parity of $\mathbf{T}_{JLM}$ is determined by that of the scalar spherical harmonics $Y_L^{M+\mu}$, which is $(-1)^L$. Then, to ensure that the fields are eigenfunctions of the parity operator, only terms with equal parity can be combined, i.e. the solution with $L=J$ must remain separate, while those with $L=J\pm1$ can be mixed. The multipoles are classified as electric or magnetic poles according to their parity: If $\mathbf{A}$ has parity $(-1)^J$, then it is a $2^J$ pole magnetic field, and in the case of $\mathbf{A}$ with parity $(-1)^{J+1}$ it is a $2^J$ pole electric field. For this reason, the magnetic multipole vector potential can only be composed from a term with $J=L$. The vector functions used are therefore
\begin{equation}
\mathbf{T}_{LLM}=\sum_{\mu} C(1LL;-\mu,M+\mu) Y_L^{M+\mu}(\theta,\phi)\boldsymbol{\xi}_{-\mu} \label{TLLM}.
\end{equation}
Then, in order to obtain the vector potential for the magnetic multipole field, $\mathbf{T}_{LLM}$ has to be multiplied by a spherical Bessel or Hankel function:
\begin{equation}\label{Aequation}
\mathbf{A}_{LM}(m)=C_L(m) z_L(kr)\mathbf{T}_{LLM}.
\end{equation}
$C_L(m)$ is a normalization factor, which is set to -1 in \cite{Rose_booklet}. Now it is very easy to obtain the electric multipoles by using (\ref{Maxwell3}) or the symmetries between the electric and magnetic fields.

\begin{figure}[htp]
\centering
\includegraphics[width=17cm]{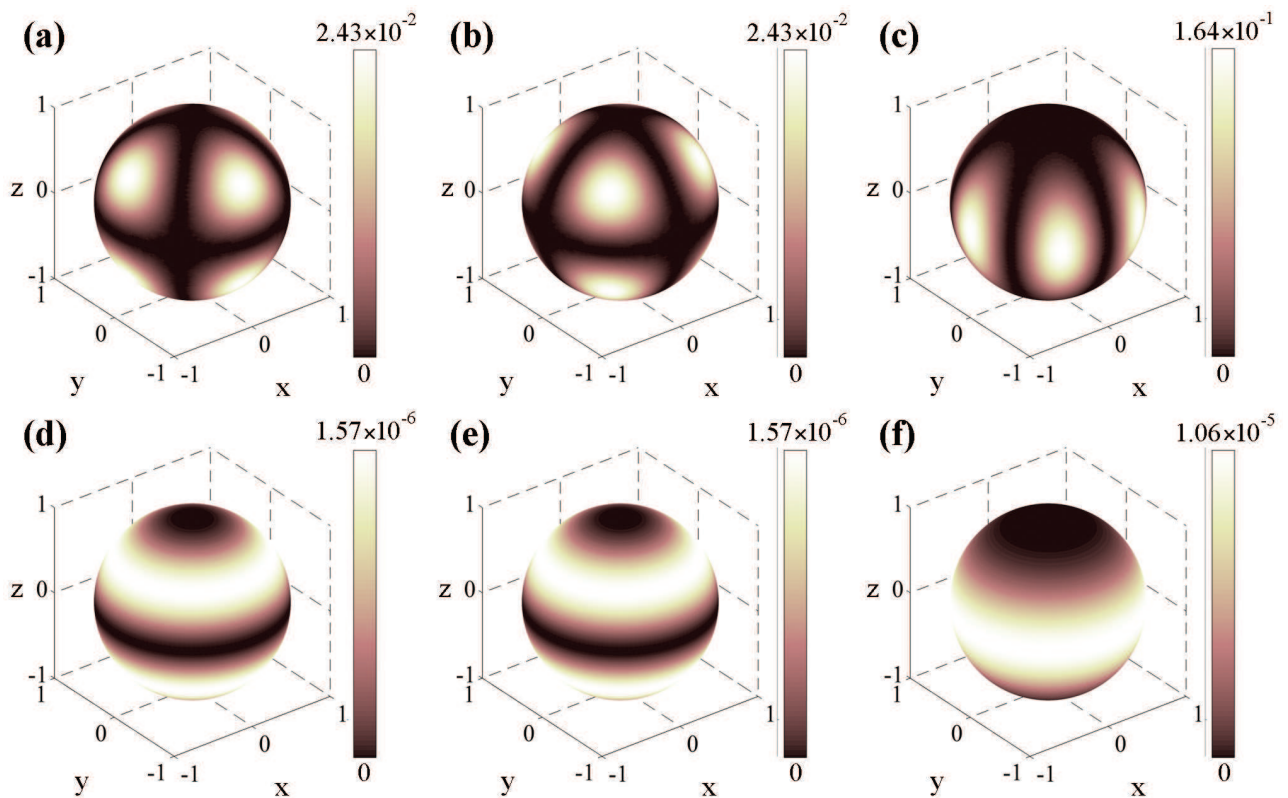}
\caption{(Colour online) Squared modulus of the x- ((a),(d)), y- ((b),(e)), and z- ((c),(f)) component of the multipole field with indices $\{3,3\}$ on a spherical surface in each formalism. Top row: $\mathbf{M}_{o33}$ from the general method. Bottom row: $\mathbf{E}_{33}(m)$ from the angular momentum method. }
\label{comparison_fig}
\end{figure}

By construction, the electromagnetic fields with this shape are eigenvectors of the angular momentum operators. Now, having followed the two derivations, it is unclear how the fields obtained by the general method are related to the ones from the angular momentum method. The final expressions for the fields from the general method (see for example (\ref{Mexample})), which are typically given in terms of spherical unit vectors, have a dissimilar presentation to those from the angular momentum method (e.g. (\ref{Aequation})). As an illustration of the difference in appearance between the two sets of solutions, the squared modulus of each component of the field $\{L,M\}=\{3,3\}$ on a spherical surface is shown in figure \ref{comparison_fig}. Comparing the corresponding field components from the two formalisms, it is evident that there is a difference in both azimuthal dependence and magnitude between the fields. In order to determine the relationship between the two sets of solutions, one could use a brute force approach to calculate the angular momentum of the fields of the general method and to calculate the conversion factors from one field to the other \cite{Bouwkamp}. Here, we will present a different method by using the properties of the angular momentum operators.

\section{Correspondence between fields}\label{Correspondence}

\subsection{Properties of the angular momentum operator}

We will start by reviewing some of the properties of the angular momentum operator, which we will use later. The angular momentum operator is a vector differential operator that can be written in the form:
\begin{eqnarray}
\mathbf{L} & = &  -i(\mathbf{r} \times  \nabla) \label{OAM}\\ 
& = & L_x \hat{\mathbf{x}}+ L_y \hat{\mathbf{y}}+ L_z \hat{\mathbf{z}} \\
& = & -L_{+1} \boldsymbol{\xi}_{-1}+ L_z \boldsymbol{\xi}_0- L_{-1} \boldsymbol{\xi}_{+1} \label{OAM_ladder}.
\end{eqnarray}
Note that (\ref{OAM}) is very similar to the quantum mechanical angular momentum operator, normalized by $\hbar$, i.e. $\mathbf{L}=\mathbf{r} \times  \mathbf{p}/\hbar$. The linear momentum is written in its differential form $\mathbf{p}=-i \hbar \nabla$. Equation (\ref{OAM_ladder}) can be easily verified with the form of the vectors $\boldsymbol{\xi}_{\mu}$ and using the so-called ladder operators:
\begin{eqnarray}
L_{+1}& = & -\frac{1}{\sqrt{2}}(L_x + i L_y) \nonumber \\
L_{-1}& = & \frac{1}{\sqrt{2}}(L_x - i L_y).  \\
\end{eqnarray}

From the theory of angular momentum it can be shown that, when applied to spherical harmonics, the ladder operators have the following effect:
\begin{eqnarray}
L_{+1} Y_L^M&=&-\sqrt{\frac{(L-M)(L+M+1)}{2}}Y_L^{M+1} \nonumber \\
L_{-1} Y_L^M&=& \sqrt{\frac{(L+M)(L-M+1)}{2}}Y_L^{M-1} \label{ladder},
\end{eqnarray}
and indeed they can be used to obtain the relations between spherical harmonics of different orders and their explicit shape.

The final property of the angular momentum operator that we will make use of is that as the generator of rotations, the operator does not affect the radial components of the fields. This property is very easy to verify by writing the angular momentum in radial coordinates or by observing that (\ref{ladder}) operate on the spherical harmonics, which are purely angular functions (they contain no radial component).

\subsection{Rewriting the general method}

Now we can turn to the general method again and observe that the fields $\mathbf{M}$ can be rewritten in the following form:
\begin{equation}
\mathbf{M}_{oML}=\nabla \times (\mathbf{r}\psi_{oML})=-\mathbf{r} \times (\nabla\psi_{oML})=-i \mathbf{L}\psi_{oML}, \label{M_OAM}
\end{equation}
and similarly for the even fields, i.e. $\mathbf{M}_{eML}=-i \mathbf{L}\psi_{eML}$. The second equality in (\ref{M_OAM}) can be obtained from the commutation relations between $\mathbf{r}$ and $\mathbf{p}$ or directly by checking terms of the kind $y \partial/\partial z - z \partial/\partial y = - (\partial/\partial y \, z -  \partial/\partial z \,y)$. 

Finally, from the explicit expression of the spherical harmonics, which is given in the appendix, it can be shown that $\psi_{eML}+i\psi_{oML} \propto Y_L^M  z_L(kr)$ and $\psi_{eML}-i\psi_{oML} \propto Y_L^{-M} z_L(kr)$. Using this and the properties of the angular momentum operator given above, we can combine the fields $\mathbf{M}$ in the following way:

\begin{eqnarray} 
&& \mathbf{M}_{eML}\pm i\mathbf{M}_{oML}=  -i  (\mp 1)^{(M)} \sqrt{\frac{4\pi}{2L+1} \frac{(L+M)!}{(L-M)!}} z_L(kr) \mathbf{L} Y_L^{\pm M} (\theta,\phi) \nonumber\\
&=&-i z_L(kr) (\mp 1)^{(M)} \sqrt{\frac{4\pi}{2L+1} \frac{(L + M)!}{(L-M)!}}\nonumber \\
&&(-L_{+1} \boldsymbol{\xi}_{-1}+ L_z \boldsymbol{\xi}_0- L_{-1} \boldsymbol{\xi}_{+1})Y_L^{\pm M}(\theta,\phi) \label{M_inOAM},
\end{eqnarray}
which now can be cast as a sum of different spherical harmonics by using (\ref{ladder}):
\begin{eqnarray}
\fl \mathbf{M}_{eML}\pm i\mathbf{M}_{oML}&=&-i z_L(kr)  (\mp 1)^M \sqrt{\frac{4\pi}{2L+1} \frac{(L+M)!}{(L - M)!}} \nonumber \\
&&\left[ \sqrt{\frac{(L\mp M)(L\pm M+1)}{2}}Y_L^{\pm M+1}(\theta,\phi) \boldsymbol{\xi}_{-1}  \pm M Y_L^{\pm M}(\theta,\phi) \boldsymbol{\xi}_0 \right. \nonumber \\
&&\left. - \sqrt{\frac{(L\pm M)(L \mp M+1)}{2}} Y_L^{\pm M-1}(\theta,\phi) \boldsymbol{\xi}_{+1} \right] . \label{M_inY} 
\end{eqnarray}

Thus, we have been able to easily write the $\mathbf{M}$ fields of the general method as a superposition of products of angular momentum eigenstates, similarly as in the angular momentum method. Now it is just a matter of comparing the factors in (\ref{M_inY}), obtained through the use of ladder operators, with (\ref{TLLM}) and (\ref{CG coefficients1})--(\ref{CG coefficients3}) and noting that they are the same, except for a common factor \footnote{That they are the same is not a lucky coincidence. The Clebsch-Gordan coefficients are precisely derived by using the angular momentum ladder operators.}, i.e. 
\begin{equation}
\mathbf{M}_{eML}\pm i\mathbf{M}_{oML}=i (\mp 1)^M \sqrt{\frac{L(L+1) 4 \pi}{(2L+1)} \frac{(L + M)!}{(L - M)!}} z_L(kr) \mathbf{T}_{LL\pm M}, \label{M-conversion}
\end{equation}
which gives us the desired relationship between the two methods. Let us consider (\ref{M-conversion}) in more detail. The constant factor on the right hand side is due to the difference in magnitude between the two sets of solutions. Based on properties of $\mathbf{T}_{LL\pm M}$ established in section \ref{ang_mom_method_derivation}, (\ref{M-conversion}) clearly shows that $\mathbf{M}_{eML}\pm i\mathbf{M}_{oML}$ is also an eigenfunction of the operators $\mathbf{J}^2$, $J_z$, and P. Conversely, note that the modes resulting from the general method, i.e. $\{\mathbf{M}_{eML},\mathbf{M}_{oML},\mathbf{N}_{eML},\mathbf{N}_{oML}\}$ are eigenvectors of $\mathbf{J}^2$ and $P$, but not of $J_z$. This is reflected by the difference in azimuthal dependence between the two sets of solutions, visible in figure \ref{comparison_fig}, which is accounted for by the use a linear combination of $\mathbf{M}$ fields in (\ref{M-conversion}). Note that the need for this linear combination stems from the particular choice of $\phi$-dependent functions during the solution of the scalar Helmholtz equation in the derivation of the fields from the general method. If complex exponential functions were chosen instead of the sinusoidal functions, the fields derived by the two methods would be directly proportional to each other.

We could similarly derive another relation for the $\mathbf{N}$ fields, but it is easier to use that $\mathbf{N}=(\nabla \times \mathbf{M})/k$ and $\mathbf{H}=-i(\nabla \times \mathbf{E})/k$ \footnote{In \cite{Rose_book} and \cite{Rose_booklet} Gaussian units are used.}, as will be evident from the conversion tables below.

\subsection{Conversion tables}\label{Conversion}

We can now present the whole conversion tables between the electric and magnetic multipoles from Rose's books (denoted by (e) and (m), respectively) and the $\mathbf{M}$ and $\mathbf{N}$ fields from the general method used in Mie theory. It is important to emphasize that $L(L+1)$ and $M$ represent the eigenvalues of the square and z-component of the total angular momentum operator, $\mathbf{J}$.  
The vector potential can then be written as:
\begin{equation}  \label{vecpot}\fl
\begin{array}{ccl}
 \mathbf{A}_{LM}(m) &=& \frac{i (-1)^{M}}{(\mathrm{sign}(M))^M \sqrt{L(L+1)}} \sqrt{\frac{2L+1}{4\pi} \frac{(L-\left|M\right|)!}{(L+\left|M\right|)!}}(\mathbf{M}_{e\left|M\right| L}+ i~ \mathrm{sign}(M)\mathbf{M}_{o\left|M\right| L})  \\
 \mathbf{A}_{LM}(e) &=& \frac{(-1)^{M}}{(\mathrm{sign}(M))^M \sqrt{L(L+1)}} \sqrt{\frac{2L+1}{4\pi} \frac{(L-\left|M\right|)!}{(L+\left|M\right|)!}}(\mathbf{N}_{e\left|M\right| L}+i~ \mathrm{sign}(M)\mathbf{N}_{o\left|M\right| L}).
\end{array}
\end{equation} 
And for the electric and magnetic fields:  
 
\begin{equation} \label{fields}\fl
\begin{array}{ccl}
\mathbf{E}_{L M}(m) &=&- \mathbf{H}_{L M}(e) = \frac{(-1)^{M+1}k}{(\mathrm{sign}(M))^M\sqrt{L(L+1)}} \sqrt{\frac{2L+1}{4\pi} \frac{(L-\left|M\right|)!}{(L+\left|M\right|)!}}(\mathbf{M}_{e \left|M\right|L}+i~ \mathrm{sign}(M) \mathbf{M}_{o \left|M\right|L}) \\
        \mathbf{H}_{L M}(m) &=& \mathbf{E}_{L M}(e) = \frac{i(-1)^M k}{(\mathrm{sign}(M))^M\sqrt{L(L+1)}} \sqrt{\frac{2L+1}{4\pi} \frac{(L-\left|M\right|)!}{(L+\left|M\right|)!}}(\mathbf{N}_{e \left|M\right|L}+i~\mathrm{sign}(M)\mathbf{N}_{o \left|M\right|L}). 
\end{array}
\end{equation}

\subsection{Application to a plane wave}\label{Application}
\begin{figure}[htp]
\centering
\includegraphics[width=14cm]{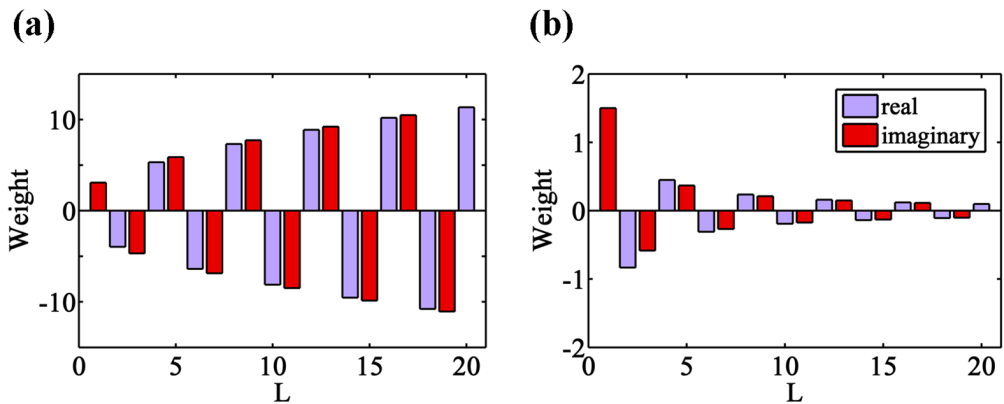}
\caption{(Colour online) Complex weights of the first basis functions in the multipole expansion of a plane wave. (a) Vector potential using Rose's notation, where weights multiply $(\mathbf{A}_{L1}(m)+i\mathbf{A}_{L1}(e) + \mathbf{A}_{L-1}(m)-i\mathbf{A}_{L-1}(e))$. (b) Electric field in the notation from \cite{Bohren}, where weights multiply $(\mathbf{M}_{o1L}-i\mathbf{N}_{e1L})$.}. 
\label{multipole_fig}
\end{figure}

As previously mentioned, the plane wave solution is very commonly used in electromagnetism. A physical problem that ties together the use of multipolar fields and plane waves is Mie scattering. The incident field is a plane wave, but due to the symmetry of the spherical scatterer it is convenient to treat the problem using multipole expansions, which necessitates the decomposition of the incident plane wave in terms of the multipolar fields. In the following, we demonstrate the use of the above conversion tables for the case of a plane wave. To convert between the two representations, we can use (\ref{vecpot}) on the plane wave decomposition given in \cite{Rose_book} to obtain the corresponding expansion in terms of $\{\mathbf{M}_{eML}, \mathbf{M}_{oML}, \mathbf{N}_{eML}, \mathbf{N}_{oML}\}$. In \cite{Rose_book}, the plane wave expansion is given as
\begin{equation}
 \mathbf{A}_{pw}=\sqrt{2\pi} \sum_{L=1}^{\infty} i^L\sqrt{(2L+1)}(\mathbf{A}_{Lp}(m)+ip\mathbf{A}_{Lp}(e)),
\end{equation}
where the vector potential for a plane wave is $\mathbf{A}_{pw}=\mathbf{u_p}e^{ikz}$ and $\mathbf{u_p}=\frac{1}{\sqrt{2}}(\mathbf{\hat{x}}+ip\mathbf{\hat{y}})$, $p=\pm 1$. Choosing $\frac{1}{\sqrt{2}}(\mathbf{u_1}+\mathbf{u_{-1}})$ to obtain a plane wave with polarization in the x direction, and applying (\ref{vecpot}) yields
\begin{eqnarray}
  \mathbf{A}_{pw} &=& \sqrt{\pi} \sum_{L=1}^{\infty} i^L\sqrt{(2L+1)}(\mathbf{A}_{L1}(m)+i\mathbf{A}_{L1}(e) + \mathbf{A}_{L-1}(m)-i\mathbf{A}_{L-1}(e))\nonumber  \\
 \fl &=&\sqrt{\pi}  \sum_{L=1}^{\infty} i^L\sqrt{(2L+1)} \frac{1}{\sqrt{L(L+1)}}\sqrt{\frac{2L+1}{4\pi} \frac{1}{(L+1)L}}\nonumber \\ &&\left[-i(\mathbf{M}_{e1L}+i\mathbf{M}_{o1L})  -i(\mathbf{N}_{e1L}+i\mathbf{N}_{o1L}) +i(\mathbf{M}_{e1L}-i\mathbf{M}_{o1L})-i(\mathbf{N}_{e1L}-i\mathbf{N}_{o1L})
\right]\nonumber  \\
\fl  &=&\sum_{L=1}^{\infty} i^L \frac{2L+1}{L(L+1)}(\mathbf{M}_{o1L}-i\mathbf{N}_{e1L}). 
 \end{eqnarray} 
 The difference between these expansions is illustrated in figure \ref{multipole_fig}, which depicts the weights of the first few basis functions with index $\{L,\pm 1\}$ in the representation from the angular momentum method, and $\{1,L\}$ in that from the general method. \\
 
\section{Conclusion}\label{Conclusion}
In this article we have compared different notations of the expressions for multipolar modes in electromagnetism. We have exposed their symmetry properties, even in the cases where they were hidden in the formalism. As a result, we have given compact explicit expressions that relate the different notations. All this has been possible by making use of the angular momentum operator and its properties. The relation between the multipolar electromagnetic modes and the angular momentum is typically overlooked in Mie theory, which may lead to some confusion. By presenting some of the most commonly used expressions for multipolar fields on equal footing, we hope to have lifted this possible confusion. 

\ack \label{Acknowledgements}
This work was supported under the Australian Research Council's Discovery Projects funding scheme (DP110103697). N. T. would also like to thank Dr. Konstantin Momot for useful discussions.

\appendix
\section*{Appendix}\label{Definitions}
\setcounter{section}{1}

For the convenience of the reader, here we will write the expressions of some of the functions we have used above.\\

\emph{Associated Legendre functions}\\
The associated Legendre functions used in \cite{Bohren} are defined as follows:
\begin{equation}
P_L^M(x)=  \frac{(1-x^2)^{M/2}}{2^L L!} \frac{d^{M+L}}{dx^{M+L}}(x^2-1)^L. \nonumber
\end{equation}
The form used in spherical harmonics has $x=cos(\theta)$, which is\\
\begin{equation}
P_L^M(\cos(\theta))=  \frac{(\sin(\theta))^{M}}{2^L L!} \frac{d^{M+L}}{d\cos(\theta)^{M+L}}(\cos^2(\theta)-1)^L. \label{associated legendre}
\end{equation}
This can be seen from page 90 of \cite{Bohren}, which provides the definition in terms of the ordinary Legendre polynomial. However, another common definition of the associated Legendre polynomial, e.g. see \cite{Jackson}, has an additional factor of $(-1)^M$. Therefore, care must be taken when comparing different references.\\

\emph{Spherical harmonics}\\
The most common definition of spherical harmonics, also used in \cite{Rose_book} (p. 241), is
\begin{equation}\fl
Y_L^M(\theta,\phi) =   \sqrt{\frac{2L+1}{4\pi} \frac{(L-M)!}{(L+M)!}} e^{iM\phi} \frac{(-1)^M}{2^L L!}(\sin(\theta))^M \frac{d^{L+M}}{d\cos(\theta)^{L+M}}(\cos^2(\theta)-1)^L.
\end{equation}

Using the definition of associated Legendre polynomials given by (\ref{associated legendre}), this means that 
\begin{equation}
Y_L^M(\theta,\phi)  = (-1)^M \sqrt{\frac{2L+1}{4\pi} \frac{(L-M)!}{(L+M)!}} P_L^M(\cos(\theta))e^{iM\phi}. 
\end{equation}

Based on this, 
\begin{eqnarray}\fl
 \cos(M\phi)P_L^M(\cos(\theta))+i \sin(M\phi)P_L^M(\cos(\theta)) &=& e^{iM\phi}P_L^M(\cos(\theta))\nonumber \\
&=& (-1)^M\sqrt{\frac{4\pi}{2L+1} \frac{(L+M)!}{(L-M)!}}Y_L^M(\theta,\phi) \\
\fl \cos(M\phi)P_L^M(\cos(\theta))-i \sin(M\phi)P_L^M(\cos(\theta)) &=& e^{-iM\phi}P_L^M(\cos(\theta)) \nonumber \\
&=&e^{-iM\phi}P_L^{-M}(\cos(\theta))(-1)^M \frac{(L+M)!}{(L-M)!} \nonumber \\
&=&  \sqrt{\frac{4\pi}{2L+1} \frac{(L+M)!}{(L-M)!}}Y_L^{-M}(\theta,\phi).
\end{eqnarray}\\

\emph{Spherical Bessel functions and Hankel functions}\\
The two spherical Bessel functions relevant for multipole expansions are defined in terms of common Bessel functions as
\begin{eqnarray}
j_L(kr)&=&\sqrt{\frac{\pi}{2kr}}J_{L+1/2}(kr) \\
y_L(kr)&=&\sqrt{\frac{\pi}{2kr}}Y_{L+1/2}(kr) .
\end{eqnarray}
Their linear combination is used to construct the spherical Hankel functions
\begin{eqnarray}
h_L^{(1)}(kr)&=&j_L(kr)+iy_L(kr) \\
h_L^{(2)}(kr)&=&j_L(kr)-iy_L(kr). 
\end{eqnarray}\\

\emph{Clebsch-Gordan coefficients}\\
\begin{eqnarray} 
C(1LL;-1,M+1)&=&-\sqrt{\frac{(L-M)(L+M+1)}{2L(L+1)}} \label{CG coefficients1} \\
C(1LL;0,M)&=& -\frac{M}{\sqrt{L(L+1)}}  \label{CG coefficients2} \\
C(1LL;1,M-1)&=& \sqrt{\frac{(L+M)(L-M+1)}{2L(L+1)}} \label{CG coefficients3}
\end{eqnarray}\\

\section*{References}
\bibliographystyle{unsrt}
\bibliography{multipole2}
\end{document}